\documentclass[manuscript]{aastex}
\usepackage{natbib}
\usepackage{color}
\shorttitle{Frerequencies of Magnetic Activities on Supergranule Boundary II}
\shortauthors{Y. Iida et al.}

\begin{document}

\title{Detection of Flux Emergence, Splitting, Merging, and Cancellation of Network Fields. II Apparent Unipolar Flux Change and Cancellation}

\author{Y. Iida}
\affil{Japan Aerospace Exploration Agency, Yoshinodai, Cyuo-ku, Sagamihara, Kanagawa, 252-5210, Japan}

\author{H. J. Hagenaar}
\affil{Lockheed Martin Advanced Technology Center, Org. ADBS, Building 252, 3251 Hanover Street, Palo Alto, CA 94304}

\and

\author{T. Yokoyama}
\affil{Department of Earth and Planetary Science, University of Tokyo, Hongo, Bunkyo-ku, Tokyo, 113-0033, Japan}

\begin{abstract}
In this second paper in the series, we investigate occurrence frequencies of apprent unipolar processes, cancellation, and emergence of patch structures in quiet regions.
Apparent unipolar events are considerably more frequent than cancellation and emergence as per our definition, which is consistent with \cite{lam2013}.
Furthermore, we investigate the frequency distributions of changes in flux during apparent unipolar processes are and found that they concentrate around the detection limit of the analysis.
Combining these findings with the results of our previous paper, \cite{iid2012}, that merging and splitting are more dominant than emergence and cancellation, these results support the understanding that apparent unipolar processes are actually interactions with and among patches below the detection limit and that there still are numerous flux interactions between the flux range in this analysis and below the detection limit.
We also investigate occurrence frequency distributions of flux decrease during cancellation.
We found a relatively strong dependence, $2.48 \pm 0.26$ as a power-law index.
This strong dependence on flux is consistent with the model, which is suggested in the previous paper.
\end{abstract}


\section{INTRODUCTION}

An understanding of magnetic structures on the solar surface is important because they function not only as a method of energy storage for various solar surface activities but also as a practical example of a magneto-convection system on the stellar surface.
Although some researches have been conducted from the perspectives of both observation and numerical simulation, many questions still remain.

One such question is the frequency distribution of the flux in the patches \citep{wan1995,sch1997,hag1999,hag2001,par2009,zha2010,iid2012,shio2012}, which is used in this series of the papers as a word corresponding to the magnetic elements, magnetic features, or flux concentrations in the literatures.
In an important discovery, \cite{par2009} reported a power-law distribution with an index of $1.85\pm0.14$ in a flux range of $2 \times 10^{17}$ Mx - $10^{23}$ Mx, namely from the observational limit to the active region.
Several subsequent papers also report similar results within a margin of error \citep{zha2010, iid2012, shio2012, zho2013}.

Despite the accumulation of observational results, the formation and maintenance mechanism of this distribution are still unclear.
\cite{par2009} suggested two possible interpretations.
One is the production of such a power-law distribution in the deeper layer of the convection zone.
This means that the flux supply from the deeper layer to the surface directly forms in the observed power-law distribution.
The other is the production by the surface processes of patches, meaning that the flux supplied from the deeper layer is deformed to the power-law distribution by merging, splitting, and cancellation on the surface.
These interpretations can be discriminated by comparing the timescales of the magnetic flux supply and surface processes.
However, despite the importance of delving further into these findings, such an investigation by counting the processes manually is difficult because such occur at a very high frequency on the solar surface.

Therefore, we developed an auto-tracking algorithm of patch structures and surface processes in the previous paper (Iida et al., 2012; hereafter Paper I). 
Note that not only the algorithm but also the stable high-resolution data acquired by the ${\it Hinode}$ allows us to perform a statistical investigation.
The occurrence frequencies of merging and splitting are found to be considerably higher than those of emergence and cancellation. 
This result is equivalent with that the reprocessing time scales of merging and splitting are considerably shorter than those of flux supply from a deeper layer, i.e., the observed distribution of flux is nontrivially affected by the surface processes.

Paper I focused only on merging and splitting because the temporal duration of the data was not enough for a statistical investigation of the other processes, i.e. emergence and cancellation.
These dynamic processes are related to various types of energetic events.
They are also considered important for the understanding of flux transport, in that they represent flux supply and removal from the solar surface \citep{liv1985, mar1985}.

Flux emergence, defined as the appearance and separation of opposite polarities, is thought to be the origin of magnetic flux on the solar surface.
It is a highly visible phenomenon with its distinct features in magnetogram images.
The separating velocity of opposite polarities reaches up to $4$km s$^{-1}$, which is significantly larger than the typical patch motion at $\thicksim 1$km s$^{-1}$ \citep{ots2007}.
There are many studies investigating the frequency distribution of the flux of emergence \citep{har1975, har1993,tit2000,hag2001,ots2011,tho2011}.
One of the most interesting questions is which scale is dominant in flux supply.
\cite{tho2011} obtained a steep power-law form with an index of $\thicksim -2.7$ by combining their results in quiet regions based on their feature-tracking and those from previous studies.
This power-law index, which is steeper than $-2$, suggests that smaller-scale flux emergence dominates the flux supply from below the photosphere.

A cancellation event is defined as a converging motion and subsequent disappearance of opposite polarities and is a relatively less distinct phenomenon in magnetogram images, with a typical velocity of less than $1$km s$^{-1}$ \citep{cha2002,par2009}.
It is considered a major process of flux removal process from the solar surface and is also related to various solar features, e.g., filament formation, flares, and EUV bright point \citep{mar1998,mart2001}. 
Two scenarios describing this process, U-loop emergence and $\Omega$-loop submergence, are suggested \citep{zwa1985}. 
Investigations of Doppler velocity, upflow or downflow, and the horizontal field direction, connecting positive to negative or negative to positive, are critical to discriminate between these scenarios.
Some recent papers report downflow structures and a normal connection in the horizontal field \citep{cha2004b,iid2010,cha2010} but are still inconclusive \citep{kub2010,wel2013}.
From the view of flux maintenance, the flux dependence of cancellation still remains uninvestigated despite its importance \citep{iid2012}. 

Besides emergence and cancellation, apparent unipolar appearances and disappearances occur without any interactions with surrounding patches.
\cite{lam2008} first investigated such unipolar appearance events statistically with the magnetograms obtained by Michelson Doppler Imager on board {\it Solar and Heliospheric Observatory}.
\cite{lam2010} extended the research and also used magnetograms by Solar Optical Telescope on board the {\it Hinode} satellite. 
They investigated an ensemble of averages of magnetic flux profiles around such unipolar appearance events and found a surrounding plateau that has the same polarity as the detected features. 
This means that there are magnetic fields undetected by the auto-tracking technique, which supports the idea that these appearances are actually a coalescence of undetected patches.
Further, \cite{lam2013} investigated the birth and death of magnetic patches. 
They determined that unipolar events are frequent dominant processes during the partial lifetime of patch structures, i.e., the expected lifetime of patches by a particular process, calculated from its occurrence frequency. 
This dominance means that they are also dominant in apparent flux removal, which is consistent with our results in Paper I.

Accordingly, to understand flux maintenance on the solar surface, the second paper in the series investigates the patch interactions accompanying changes in flux in patches, namely the apparent unipolar process, cancellation, and emergence, with a focus on the flux dependence of their occurrence.
In Sections 2 and 3, observational data and definitions of flux change events, emergence, and cancellation in our analysis are described.
Results are shown in Section 4. 
Finally, these results are discussed and one plausible picture of flux maintenance in quiet regions is presented in Section 5.

\section{OBSERVATION}
Two data sets of magnetograms taken by a Solar Optical Telescope (SOT) / Narrowband Filter Imager (NFI) onboard the {\it Hinode} satellite \citep{kos2007, ich2008, tsu2008, sue2008}, are used.
One is the data set also used in Paper I (hereafter Data1).
To increase number of detected cancellations and emergences there, another data set of magnetograms is used, with the longest duration hitherto obtained by the {\it Hinode} satellite (hereafter Data2).
For a detailed description of Data1, refer to Section 2 in Paper I. 

The observational period of Data2 is from 10:24UT on December 30, 2008 to 5:37UT on January 5, 2009, a total duration of 115h 13min.
The {\it Hinode} satellite is usually pointed to the disk center for the synoptic observation. 
During this period, all the synoptic observaton are cancelled and the data is uninterrupted for the long duration.
The SOT tracks a quiet region near the solar equator.
The center of the SOT view moves from $(-520.9'',-9.3'')$ to $(695.9'',-6.9'')$ in the solar coordinate.
SOT/NFI obtains circular polarization images, namely Stokes-V/I, at one wavelength shifted by +140m$\AA$ from the Na$_{\rm I}$ D$_1$ resonance line at 5896$\AA$.
The temporal cadence is $5$ min.
The original pixel scale of the CCD camera is $0.08''$ for this band.
The data are summed by 2 pixels in both the x- and y-directions onboard and are recorded with a spatial binsize of $0.16''$.
The total view is $704$ pixels $\times$ $704$ pixels, which is equivalent to $112.6'' \times 112.6''$.
Note that, because the center of the view is gradually shifted because of the long duration of the observation, we extract the field of view contained through throughout the entire observational period, which is explained in detail later.
The final analyzed view measures $94.9'' \times 91.4''$ after this extraction.
An example of a magnetogram from Data2 is shown in Figure 1.
We can observe some network cells in the field of view, as with Data1.
The total number of retrieved images is 1642 after removing the frame with missing data because of a readout error of CCD camera.
The cadence of analyzed data is $\thicksim$5 min though it occasionally 10 min due to this removal.
Table 1 shows a summary of the observational parameters in Data1 and Data2.

Before detection and tracking of patch structures, the cleaned-up magnetogram images must be prepared.
Such preparations are performed in the same manner as for Data1 in Paper I: dark current subtraction and flat fielding by the ${\rm fg\_prep}$ procedure in the SolarSoftWare package; removal of the columnwise offset of SOT/Filtergram (FG) data by subtracting the average in each column \citep{lam2010}; conversion of each pixel from a circular polarization signal to magnetic flux by fitting a scatter plot of the circular polarization signal, obtained by the SOT/FG, and magnetic flux, obtained by Solar and Heliospheric Observatory ({\it SOHO})/ Michelson Doppler Imager (MDI); spatial and temporal averaging to smooth the observed signal; and de-rotation to a certain time, for which we use 20:35UT on January 1 for Data2 when the observed region is near the disk center. 

In addition to these processes, two more corrections are needed due to the long observational period of Data2.
First, the observed line-of-sight magnetic field is projected onto the field perpendicular to the solar surface, simply assuming that the magnetic field is always perpendicular to the solar surface.
The vertical magnetic flux density in each pixel is calculated as
\begin{equation}
B_{\rm ver} = \frac{B_{\rm LOS}}{\cos \theta} 
\end{equation}
where $\theta$ is the heliocentric angle of each pixel.

Next, the gradual shift in the field of view that occurs due to the small difference between the tracking curve of the satellite and the rotation of the Sun is removed by making correlations between consecutive magnetograms
There is one other change of the observational field of view in the long period observation by {\it Hinode}/SOT. 
The difference between the tracking curve and rotation accumulates, and when the position of the tip-tilt mirror hits the stroke limit, it returns to its original position.
Thus, there is one sudden change of field of view in Data2.
We limit our analysis only to the field of view contained throughout the entire observational period.
The final analyzed field of view measures $94.9'' \times 91.4''$, which is slightly smaller than the original field of view of $112.6'' \times 112.6''$.

Figure 2 shows the time series of the total flux contained in the analyzed field of view.
Though we cannot observe any active regions during this observational period, there are some periods during which both polarities simultaneously increase; these are periods of relatively large flux emergence.
Quantitatively, the total flux averaged over the observational period are $1.70 \times 10^{20}$ Mx, $1.54 \times 10^{20}$ Mx, and $3.24 \times 10^{20}$ Mx for positive, negative, and both polarities, respectively.
These correspond to $3.76$ Mx cm$^{-2}$, $3.42$ Mx cm$^{-2}$, and $7.16$ Mx cm$^{-2}$.
Because these averaged flux densities are not considerably larger than those in Data1, $2.53$ Mx cm$^{-2}$ and $3.60$ Mx cm$^{-2}$, this region is thought to be quiet region.
In case of flux balance, the difference in the flux of positive and negative polarities is, on average, less than 9$\%$.

\section{DETECTION OF UNIPOLAR FLUX CHANGE, CANCELLATION, AND EMERGENCE}
We use the same methods to track the patch structures and detect merging and splitting events as in Paper I. 
Clumping method is used in the recognition of patches.
The signal and size threshold are set as two sigma of the histogram of magnetic flux density in each image and 81 pixels respectively.
We employ relatively large value for the size threshold with an aim to pick up only patches on the network boundary.
Checking spatial overlaps does the tracking between the consecutive images.
Merging and splitting are also determined by spatial overlaps.
Refer to Section 3 in Paper I for more detailed explanation of these methods. 
Note that we need a margin of five extra pixels for tracking.
This is because the time interval of the consecutive magnetogram is 5 minutes and the expected travel distance is larger than the scale of 1 pixel in the {\it Hinode}/SOT magnetogram.
Our method of detecting unipolar flux change events, cancellation, and emergence are described below. 

First, significant flux change events, excluding merging and splitting, are detected in each tracked patch.
A significant flux change event is defined as a flux increase or decrease that has a duration and flux change rate exceeding certain threshold values.
Because this step occurs after the detection of merging and splitting, i.e., we already know which patches experience merging and splitting, and when, flux changes caused by merging and splitting can be excluded.
We subtract the flux of other merged parent patches after the occurrence of merging and repeat the same step for splitting. 
Based on this concept, a corrected time series of flux in a certain patch, $\phi_{\rm COR}$, is calculated as 
\begin{equation}
\phi_{\rm COR}(t) = \phi(t)
- \sum_{j=1}^{N_{\rm mrg}} \sum_{k=1}^{N_{\rm mrg}^j} \phi_{\rm DAU, k}^j(t_{\rm mrg, j}) H(t-t_{\rm mrg, j})   
+ \sum_{j=1}^{N_{\rm splt}} \sum_{k=1}^{N_{\rm splt}^j} \phi_{\rm PAR, k}^j(t_{\rm splt, j}) H(t-t_{\rm splt, j})   
\end{equation} 
where $\phi(t)$ is the time series of the flux of the considered patch; $\phi^i_{\rm DAU, j}$ and $\phi^i_{\rm PAR, j}$ are the time series of the flux of the $j$-th daughter patches for the $i$-th merging and the $j$-th parent patches for $i$-th splitting, respectively; $N_{\rm mrg}$ and $N_{\rm splt}$ are the number of merging and splitting events, respectively; $N_{\rm mrg}^i$ and $N_{\rm splt}^i$ are the number of daughter and parent patches, respectively, excluding the considered patch in the $i$-th merging and splitting; $t_{\rm mrg, i}$ and $t_{\rm splt, i}$ are the time intervals of the considered merging and splitting occurrences, respectively; and $H(t)$ is the Heaviside step function.
The second term in corresponds to the total flux of daughter patches in the merging events and the third term does that of parent patches in the splitting events.

To pick up flux change events, we set duration and flux change rate thresholds based on the typical values for cancellation and emergence events.
Cancellations in a network field have a time scale of a few tens of minutes \citep{cha2004b,iid2010}, whereas flux emergence has a shorter time scale of less than $10$ min \citep{ots2007}. 
Because we set as short a threshold for temporal duration as possible to detect emergences, the threshold for temporal duration is set as 5 min in Data1 and 10 min in Data2.
Note that the time interval in Data2 is 5 min.
The threshold for the flux change rate is derived from the typical flux change rate for cancellation.
This is because cancellation is thought to be a more moderate event in terms of flux change rate than flux emergence.
There are some reports of flux change rates of cancellation events in network fields \citep{cha2002, park2009}.
\cite{cha2002} reported a cancellation flux change rate of $3.5 \times 10^{18}$ Mx hr$^{-1}$ from their analysis of magnetograms obtained by ${\it SOHO}$/MDI. 
We use a slightly smaller value, $10^{18}$ Mx hr$^{-1}$, as the threshold for flux change rates to detect as many cancellations as possible. 
Applying these thresholds and averaging flux over three data points in the temporal direction to obtain a smoothed data, we detect significant flux change events, excluding merging and splitting. 

Figure 3 shows an example of the time evolution of flux in one tracked patch in Data1.
The dotted line represents the flux before the removal of flux changes from merging and splitting, $\phi(t)$.
The solid line represents the flux after the removal of flux changes, namely $\phi_{\rm COR}(t)$.
After the removal, three large flux change events, indicated by shaded area, can be seen in the solid line in Figure 3.
These flux changes are detected as significant flux change events.

Next, we make pairs of the flux change events among positive and negative polarities within the certain spatial and temporal overlaps.
We set the same spatial threshold for cancellation as that for tracking patches, 1 pixel in Data1 and 5 pixels in Data2, because the typical proper velocity of canceling patches is less than 1 km s$^{-1}$ \citep{cha2002, park2009}.
Remind that the temporal cadence of Data2 is 5 minutes and one fifth of that of Data1.
On the other hand, an emerging bipole is known to have a larger velocity than that of typical patch motion \citep{ots2007}.
We set the threshold for emergence at 4km s$^{-1}$, which corresponds to 4 pixels in Data1 and 20 pixels in Data2.
In the case of temporal overlapping, three data points in the temporal direction in both Data1 and Data2 are measured and averaged.
The pairing is done from the events with larger flux changes, which guarantees the uniqueness of the pairing.
After forming pairs between opposite polarities, unpaired flux change events are categorized as unipolar flux change events.

Our method can detect partial cancellations, i.e., those in which one or both the involved patches do not disappear. 
It can also detect partial emergences by the same logic. 
However, this method categorizes a hybrid event of flux increase and decrease, i.e., the simultaneous occurrence of cancellation and emergence, 
as a single cancellation, a single emergence, or a mix of uni-polar processes. 
Thus, the detected numbers for emergence and cancellation events may underestimate the actual numbers.
Another item to note is why a pair of flux increases in patch structures are defined as a flux emergence.
It may seem that a pair of flux appearances should be defined as an emerging flux; however, as will be pointed out in Figure 4, because emerging bipoles coalesce with pre-existing patch structures soon after their appearance, tracking an emerging bipole from its birth is difficult.
A final note addresses our definition of the change in flux during emergence and cancellation events.
Larger change in flux of positive or negative polarities in each event is employed because the first part of appearance and last part of disappearance cannot be followed because of size thresholds of detection.
This is also shown in Figure 4.

Based on the method described above, flux change events in each patch are detected and categorized into unipolar flux change, cancellation, and emergence events. 
Unipolar flux change events will be shown and investigated in Section 4.2. 
The present section concentrates on cancellation and emergence in this section.

Figure 4 shows examples of paired flux change events, namely emergence and cancellation.
The upper panels in Figure 4 show time evolutions of magnetograms in an emergence event.
There is a pre-existing positive patch in the north of the emerging region in the first panel.
The emerged negative polarity is small and unrecognizable under our detection conditions around its birth at 1:09UT.
It continues to grow larger, and when it becomes large enough at 1:13 UT, the appearance of the patch structure is detectable with our threshold.
On the other hand, an isolated patch structure of positive polarity cannot be detected because it merged with the large positive patch.
However, we can observe a continuous increase in the flux of the positive polarity from 1:11 to 1:18 UT.
After the appearance of both polarities, the distance between them continues to grow from 1:19 to 1:23 UT. 
It can be observed that the quiet region is filled with a magnetic field, making it very difficult to detect isolated patch structures from their birth.
Many coalesce with pre-existing patches very soon after their birth.

The lower panels in Figure 4 show a typical example of a cancellation event.
One negative patch shown by the yellow arrow, is converging with a larger positive patch in the north.
They contact at 0:49 UT and begin to cancel each other.
The negative patch continues to shrink, and it becomes smaller than our size threshold at 1:05 UT.
Subsequently, it cannot be tracked with our threshold and totally disappears below the detection limit at 1:17 UT.
On the other hand, the positive patch remains even after the cancellation because the flux of the positive patch is larger than that of the negative patch, which results in a partial cancellation.

These events show that tracking patches from their exact time of birth to their exact time of death within certain size threshold limits is impossible. 
This suggests that estimated values of the change in flux during these events are smaller than the actual values. 

\section{RESULTS}

\subsection{Frequencies of each event and frequency distribution of flux in Data2}
Table 2 summarizes the incidence of detection for both Data1 and Data2.
The number of detected processes in Data2 increases from Data1 by approximately one order of magnitude because of the long duration.
The number of detected cancellations becomes particularly large, sufficient for a statistical analysis of Data2, as we expected.
On the other hand, the number of detected emergence events is still small and not sufficient for an investigation of scale dependence. 
The occurrence of splitting and merging is roughly $7 \-- 8$ times that of cancellation, which is consistent with Paper I.
There are many unipolar events.
Furthermore, the incidence of all four unipolar processes has the same order of magnitude for each data set. 
Despite the fact that the requirement for the detection of an emergence event is simply an occurrence of flux increase of opposite polarities within a certain distance, the number of emergence events is still considerably lesser than other processes.
This definition is less strict than those used in  previous papers on emergences \citep{hag2001,tho2011}, but we employ a larger size threshold to detect network structures.

Figure 5 is the frequency distribution of the flux in Data2.
We fit the range increasing monotonically from $10^{17.7}$ Mx indicated by the green dashed line in the figure.
A slightly smaller value, $10^{17.5}$ Mx, was obtained in Data1.
By fitting the data with a power-law function of the form $n(\phi)=n_0 \, (\phi/\phi_0)^{-\gamma}$ in the range from $10^{17.7}$ Mx to $10^{19}$ Mx, we obtain $n_0 = 1.63 \times 10^{-36}$ Mx$^{-1}$ cm$^{-2}$ and $\gamma = 1.93 \pm 0.03$ with $\phi_0 = 10^{18}$ Mx.
The errors show a 1$\sigma$-error in the fitting.
Alterations in the power-law index arising from different fitting ranges are also checked.
By considering these errors, the power-law index is $\gamma=1.93\pm0.07$, which is consistent with the previous studies \citep{par2009, iid2012, zho2013}.

\subsection{Unipolar Flux Changes}

As described in the previous section, many flux change events were detected in apparently isolated patches.
We call these flux change events unipolar processes in this study and categorize them into increase, appearance, decrease, and disappearance.
A unipolar increase is a significant increase in flux without recognizable paired flux increase events of opposite polarity.
A unipolar appearance is a significant increase in flux by a newly appeared patch but without a paiered flux increase of opposite polarity patch.
The other unipolar events are defined in a similar manner.

Figure 6 shows typical examples of unipolar flux change events in Data1.
We use Data1 here because of its better temporal cadence of the magnetogram than Data2. 
Figure 6(a) is a unipolar increase event.
There is a negative patch at 1:46 UT, indicated by the blue contour around the center of the field of view (patch A).
In addition, there are two small patches without contours located to the east of patch A (patches B and C shown by the yellow arrow), which are below our detection threshold.
Patch A absorbed B and C to become larger at 2:01 UT.
Because patches B and C are below our detection limit, this event is recognized simply as an increase of flux in patch A, i.e., a unipolar increase event.

Figure 6(b) shows a unipolar appearance event.
In this case, at 2:29-2:31 UT, there are three small patches below the detection limit, which are shown by the yellow arrow. 
They converge and coalesce to one large patch structure at 2:33 UT and retain the shape of the individual patches until 2:47 UT.
This event is explained in a similar manner in \cite{lam2010}, as a coalescence among undetected patches.

Figure 6(c) shows a unipolar flux decrease event. 
One detected positive patch shown by the yellow arrow, is located near the center of view at 1:41 UT.
A hump appears in the eastern part of the patch at 1:42 UT and separates later at 1:46 UT.
The patch structure gradually grows until 1:45 UT and is finally separates at 1:46 UT.
The separated patch is below our detection limit.

Figure 6(d) shows a unipolar disappearance event.
One detected negative patch is indicated by the yellow arrow at 2:19 UT.
A dipped shape suddenly forms in the outline of the patch at 2:27 UT.
At 2:29 UT, the patch is divided into two small negative patches below our detection limit, and they separate from each other.

All the unipolar processes depicted in Figure 6 are explained as interactions with and among patches below the detection limit, which supports the conclusion reached by \cite{lam2010} that unipolar appearance is actually the coalescence of undetected patches. 

Based on the interpretation that these apparent unipolar processes are interactions among patches below the detection limit, the change in the flux during these processes should be close to the detection limit because of the higher frequency of interactions between two patches than that among more than two patches.
Figure 7 shows histograms of changes in the flux during unipolar processes.
All four histograms concentrate around the detection limit, $10^{17.5}$ Mx, which is shown by the green dotted line in the figure.
This result supports the above interpretation. 

\subsection{Frequency Dependence of Change in Flux During Cancellation}
We investigate the frequency distribution of changes in the flux during the cancellations by using Data2.
The distribution is derived from the observational results in the same manner as that of merging and splitting in Paper I, namely
\begin{equation}
\frac{\partial n_{\rm cnc}(\phi)}{\partial t} = \frac{N_{\rm cnc}^{\rm tot}(\phi)}{t^{\rm tot} \, S \, \Delta \phi},
\end{equation}
where $N_{\rm cnc}^{\rm tot}$ is the total number of cancellation events for patches with the flux  from $\phi$ to $\phi+\Delta \phi$, $t^{\rm tot}$ is the duration of the data, and $S$ is the area of the total field of view.
Figure 8 shows the resultant plot.
We can observe power-law-like dependence above the detection limit, $10^{17.7}$ Mx.
To quantify the dependence, a least-squares fitting is applied as the power-law distribution.
The fitting form employed is
\begin{equation}
\frac{\partial n_{\rm cnc}}{\partial t}= \left( \frac{\partial n}{\partial t} \right)_{0,{\rm cnc}} \left( \frac{\phi}{\phi_0} \right)^{-\gamma_{\rm cnc}},
\end{equation}
where $\left( \partial n/ \partial t \right) _{0,{\rm cnc}}$ is the reference frequency of cancellation, $\phi_0$ is the reference flux, and $\gamma_{\rm cnc}$ is the power-law index of cancellation occurrence.
We obtained $\left( \partial n / \partial t \right) _{0,{\rm cnc}}=(1.29\pm0.12) \times 10^{-41}$ Mx$^{-1}$ cm$^{-2}$ s$^{-1}$ and $\gamma_{\rm cnc}=2.48\pm0.26$ with $\phi_0 = 10^{18}$ Mx and a fitting range of $10^{17.7}$ Mx$\-- 10^{18.4}$ Mx.
The errors indicate 1$\sigma$-errors of least-squares fitting.
We also check the obtained power-law index with different bin sizes from 0.1 to 0.24.
Figure 9 shows the result of this investigation.
The range of the power-law index from $2.3$ to $2.6$ stays within the 1$\sigma$-error range in all cases; thus, we conclude that the proper power-law index obtained here is from $2.3$ to $2.6$.

Further, we apply the Kolmogorov-Smirnoff goodness-to-fit test on power-law fit of the cancellation frequency distribution.
The Kolmogorov-Smirnoff statistic, $D$, is introduced as a maximum difference of the cumulative distributions of the reference, power-law here, and the actual data.
The critical value $D_{\rm crit}$, with which we compare $D$, is calculated from number of data point and confidence level we set.  
See \cite{par2009} for more detailed explanation of the Kolmogrov-Smirnoff test.
With number of analyzed cancellations in Data2, $362$ events, and the most permissive value of the confidence level of 99$\%$, we obtain the $D_{\rm crit}=0.0857$ from the Kolmogorov-Smirnov Table.
On the other hand, $D=0.186$ is obtained from our result.
This result, $D > D_{\rm crit}$, implies that the power-law fit is not good with the confidence level of 99$\%$.

\section{DISCUSSION}

\subsection{What are Unipolar Processes?}
\cite{lam2010} first investigated the unipolar appearances and found a same polarity plateau with the same polarity as the detected features, which supports the concept that these events are actually coalescences of undetected patches.
This paper has also offered four results supporting that all detected apparent unipolar processes, disappearances as well as appearances, are interactions with and among patches below the detection limit (Figure 7).
We want to emphasize that our result shows this result in the different way from that of \cite{lam2010}.
They compared the simultaneous magnetograms with different resolutions, {\it SOHO}/MDI and {\it Hinode}/SOT.
Only the magnetogram by {\it Hindoe}/SOT is used in this paper but we can compare detected-scale and undetected-scale by the large size threshold for the patch detection.

Several statistical findings are also important.
First, the frequency of incidence of such unipolar processes is much considerably higher than that of cancellation and emergence events and is of the same order as the frequencies of merging and splitting.
Second, all the four types of unipolar processes have occurrence frequencies of the same order.
As shown in Figure 10, there are actual physical possibilities that these unipolar processes occur through merging and splitting, but not through cancellation.
From the finding that merging and splitting are dominant processes on the solar surface, which is investigated in Paper I, these four processes occur in frequencies of the same order.
Third, the frequency distributions of changes in flux during the apparent unipolar processes is concentrated around the detection limit.
This result also supports the aforementioned interpretation because the interactions in Figure 10 result in apparent unipolar flux change events that change the flux by an amount that is at most twice the detection limit.
All these characteristics support the interpretation that apparent unipolar processes are in fact interactions involving small patches.

\subsection{Comparison with the Cancellation Model Suggested in Paper I}

The frequency distribution of canceling flux is evaluated by Eq.(18) of Paper I as
\begin{equation}
\frac{\partial n^{\rm APP}_{\rm cnc}}{\partial t}=\frac{v_0 n_0^2 \phi_0}{2(\gamma -1)\sqrt{\rho}}\left(\frac{\phi}{\phi_0}\right)^{-2\gamma+1},
\end{equation}
where $v_0$ is the typical velocity of patches, $n_0$ is the reference frequency of the flux of patches, $\gamma$ is the power-law index of flux distribution of patches, and $\rho$ is the number density of network cells. 
Comparing this and Eq.(4), we obtain two relationships: 
\begin{equation}
\gamma_{\rm cnc}=2\gamma-1
\end{equation}
for the dependence on flux, and that
\begin{equation}
\left( \frac{\partial n}{\partial t} \right)_{0,{\rm cnc}}=\frac{v_0 n_0^2 \phi_0}{2(\gamma -1)\sqrt{\rho}}
\end{equation}
for the magnitude.

In Data2, comparing the observational results and this model is possible.
From the model presented here, $\gamma=1.93\pm0.07$ was calculated and $\gamma_{\rm cnc}$ was evaluated as $2.86\pm0.14$ by Eq.(6). Observational measurement yielded a similar value of $\gamma_{\rm cnc}$ as $2.48\pm0.38$. This consistency in dependence index supports our model.
In Eq.(7), we adopt $v_0=0.5 \, {\rm km \, s}^{-1}$ and $\rho=1.0 \times 10^{-19} \, {\rm cm}^{-2}$ as typical values \citep{hag1997}, yielding an amplitude of $3.6 \times 10^{-41} {\rm Mx^{-1} cm^{-2} s^{-1}}$, which is larger than our observation, $\partial n_{0,{\rm cnc}} / \partial t=(1.29\pm0.12) \times 10^{-41}$ Mx$^{-1}$ cm$^{-2}$ s$^{-1}$.
This difference is possibly because the transport of patches cannot be treated as a ballistic motion with a constant velocity. Future work will improve upon this model.

Furthermore, Paper I discussed the possibility of the recycling of flux through submergence by cancellation and re-emergence in small scales.
\cite{tho2011} reported the frequency distribution of emerging flux as $3.14 \, \times \, 10^{-14} \times \left( \phi  \, [{\rm Mx}] / 10^{16} \right)^{-2.69} \, {\rm Mx^{-1} \, cm^{-2} \, day^{-1}} = 1.52 \times 10^{-40} \times \left( \phi \, [{\rm Mx}] / 10^{18} \right)^{-2.69} \, {\rm Mx^{-1} \, cm^{-2} \, s^{-1}} $.
The power-law index is consistent with the recycling model; however, the magnitude of occurrence is different by one order of magnitude.

This contradiction is significant and flux recycling probably does not happen in the flux scale investigate in this paper.
However, there are several possibilities of flux recycling.
One possibility is that the submerged flux is fragmented after the submergence and hence the frequency distribution of re-emerged flux is different from that of cancellation.
The second possibility is that flux recycling does not happen in the network field but does in the other convective structure. 
Note that our analysis is mainly limited to the network field due to the relatively large size threshold, 81 pixels.
There is a strong downflow on the network boundary.
It is possible that such downflow forces the canceled magnetic field to penetrate deeper, which results in total submergence into the convective zone. 
Future investigation at smaller scales is important.
The last possibility is that our detection method is not as good for emergence as for cancellation.
As we mentioned, the detection of emergence is more difficult than that of cancellation due to the shorter time scale and higher velocity of emerging patches.
Actually there are several undetected emergences in Figure 7, which are pointed out.
It will be helpful to compare the results from different detection methods. 

\subsection{Flux Maintenance in Quiet Regions}

Based on the analysis performed in Paper I and this paper, we suggest a scenario of flux maintenance in quiet regions.
Figure 11 shows a schematic describing this proposed scenario.

First, after the flux is supplied from below the photosphere and active regions, magnetic patches are advected, predominantly by convection on the solar surface.
During advection, a power-law frequency distribution of the flux is formed and maintained at an index of $\thicksim 1.8$ by merging and splitting on the solar surface.
This occurs when the time scales of merging and splitting are considerably shorter than those of emergence and cancellation, which is shown in Paper I, and means that the flux distribution is formed as a power-law distribution before the encounter of opposite polarities.

Quantitative investigations of this scenario are still an open question for future works. The work of \cite{mey2011} is notable. They conducted a mathematical simulation of a magnetic carpet with the observed flux distribution of emergence and frequent splitting. As a result, they obtained a flux distribution of patches similar to observational results, e.g., a power-law distribution with an index of $1.8\--1.9$.

During and after the maintenance by merging and splitting, patches are advected by convective motion. If one assumes that cancellation is caused only by advection and is nearly independent of patches' flux, it results in a steep dependence of cancellation occurrence rate on flux, i.e., a power-law index of $\thicksim 2.6$.
This is shown in the discussion of Paper I, and the present paper confirms this speculated steep power-law index observationally, though we caution that the power-law fit to the observations is not a statistically good fit.
Note that the magnitude is not precisely consistent with this model, and it needs improvement.

Then, through the cancellation, the flux submerges below the photosphere and later rises to the solar surface again.
This re-appearance may result in a steep power-law distribution of emergence on flux because the distribution of cancellation, i.e., submergence of flux, also has a steep power-law distribution. 
Although the steep power-law distribution of cancellation is found, there is a contradiction in the magnitude between the frequency distribution of cancellation in this study and that of emergence in \cite{tho2011}. 
There are several possibilities explaining this result, the fragmentation to smaller scale in the convection zone, the convective flow effect, or such recycling does not occur on the actual solar surface. 
Further investigation is needed to understand the flux maintenance in quiet region.

\section{SUMMARY OF RESULTS}
Extending the investigation of surface magnetic processes based on the auto-tracking technique employed in Paper I, this paper has investigated apparent unipolar processes, cancellations, and emergences. 
It was found that all four apparent unipolar events are considerably more frequent than cancellation and emergence events, which is consistent with \cite{lam2010} and \cite{lam2013}, and that all four types occur in frequencies of the same order of magnitude, which supports that these processes are related to the same actual processes, i.e., merging and splitting.
This paper further investigated the frequency distribution of flux changes during apparent unipolar processes.
It was found that all distributions were concentrated around the detection limit set for the experiment, which also supports the interpretation that these processes are in fact merging and splitting with and among undetected patches. 
This interpretation also implies s physical picture of many interactions within the flux range in this analysis and below the detection limit.

As for cancellation and emergence events, the occurrence frequency distribution of decrease in flux during cancellations was investigated, but not for emergences owing to the lack of event numbers.
As expected from the discussion in Paper I, a steep power-law frequency dependence, $2.48 \pm 0.26$ as a power-law index, was found for cancellation.

From these results, we proposed one plausible picture of local flux maintenance in quiet regions, shown in Figure 11.
The important assumption of this model is that merging and splitting, not cancellation and emergence, are dominant process that is confirmed by this paper and previous studies \citep{lam2010, iid2012, lam2013}. Notably, cancellation occurs by the advection in a random direction and includes local recycling rather than actual flux supply.
However, the quantification of this plausible modeling is an open question for future works.
For a deeper understanding of flux maintenance on the solar surface, a mathematical simulation of the proposed model will allow insight into whether our picture can exist in the quantitative sense. 
From the viewpoint of dynamics, we believe that analyses of spectropolarimetric observation and RMHD numerical simulation are plausible candidates to understand what dynamic process causes this picture, e.g., frequent splitting of patches on the solar surface.

\section*{Acknowledgement}
First of all, we appreciate the referee for a lot of helpful comments and advices.
The paper is much improved with the advices.
Hinode is a Japanese mission developed and launched by ISAS/JAXA, in collaboration with the domestic partner NAOJ and international partners NASA and STFC (UK). Scientific operation of the Hinode mission is conducted by the Hinode science team at ISAS/JAXA. This team mainly consists of scientists from institutes of partner countries. Support for the post-launch operation is provided by JAXA and NAOJ, STFC, NASA, ESA, and NSC (Norway). Also the authors would like to thank Enago (www.enago.jp) for the English language review.

\bibliographystyle{apj}
\bibliography{iida}

\begin{figure}
\epsscale{0.99}
\plotone{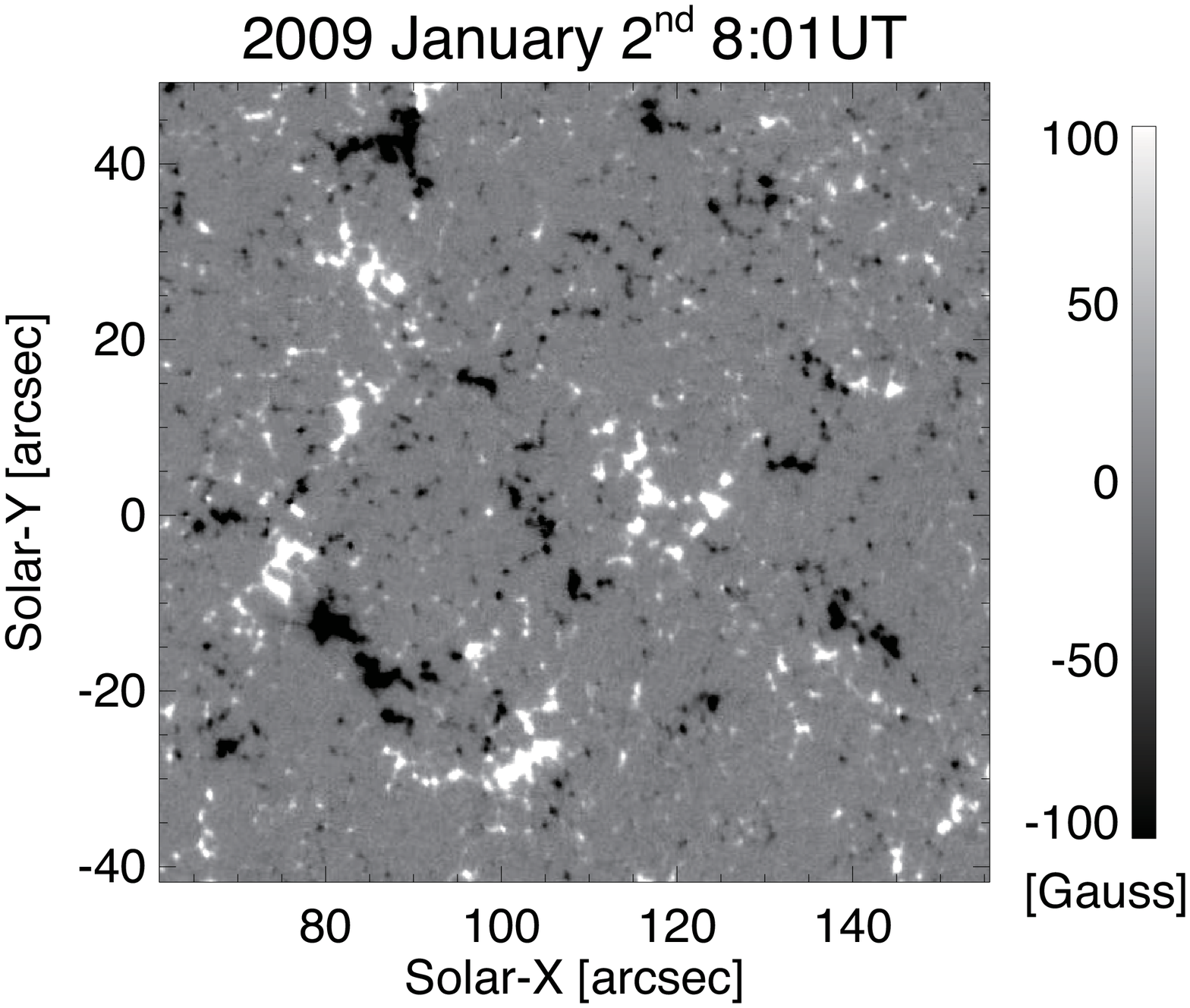}
\caption{Example of Na$_{\rm I}$ D$_1$ magnetograms of Data2.}
\end{figure}

\begin{figure}
\epsscale{0.99}
\plotone{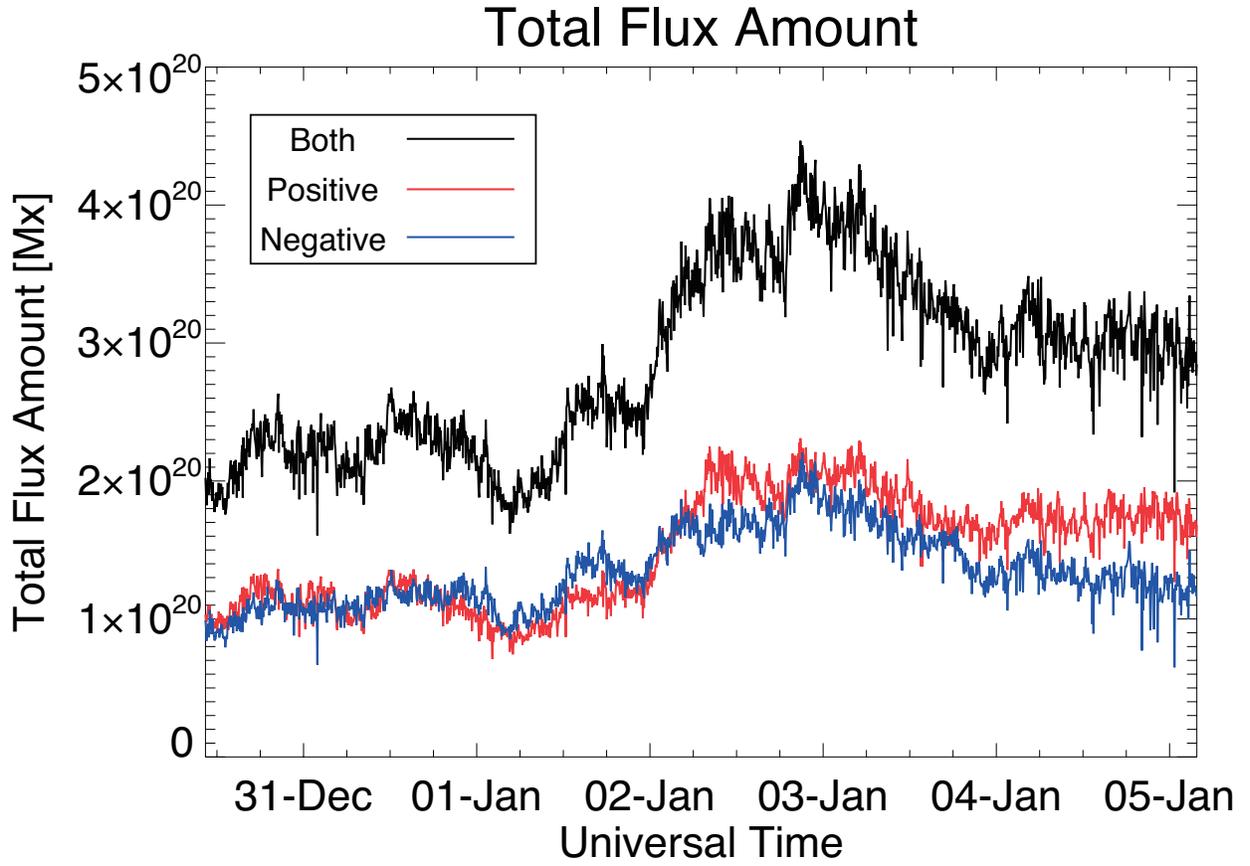}
\caption{Time series of the total flux contained in the field of view during the observational period in Data2. The red and blue lines indicate the total flux of positive and negative polarities, respectively. The black line indicates the unsigned flux.}
\end{figure}

\begin{figure}
\epsscale{0.99}
\plotone{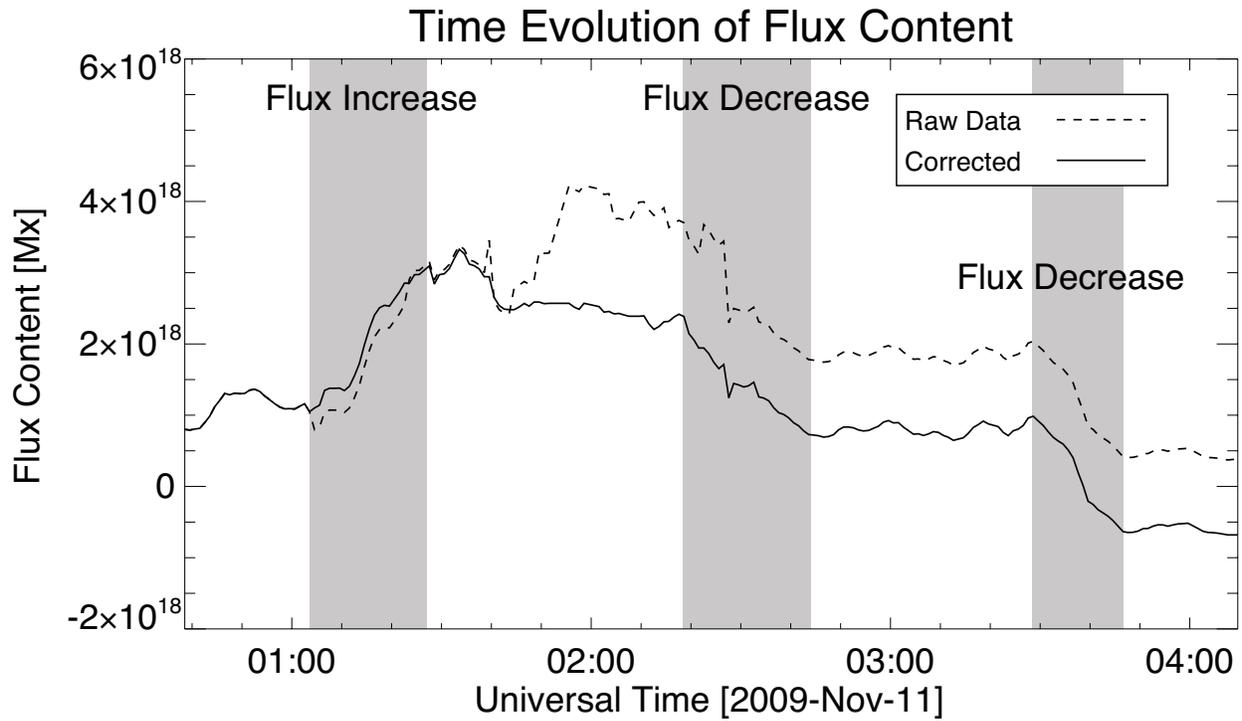}
\caption{Time series of flux and detected flux change events in a positive patch.
The dashed line indicates the actual flux, and the solid line indicates the flux after removing flux changes by merging and splitting events.
The time range for the detected flux increase and decreases are shown by the shaded areas.}
\end{figure}

\begin{figure}
\epsscale{0.99}
\plotone{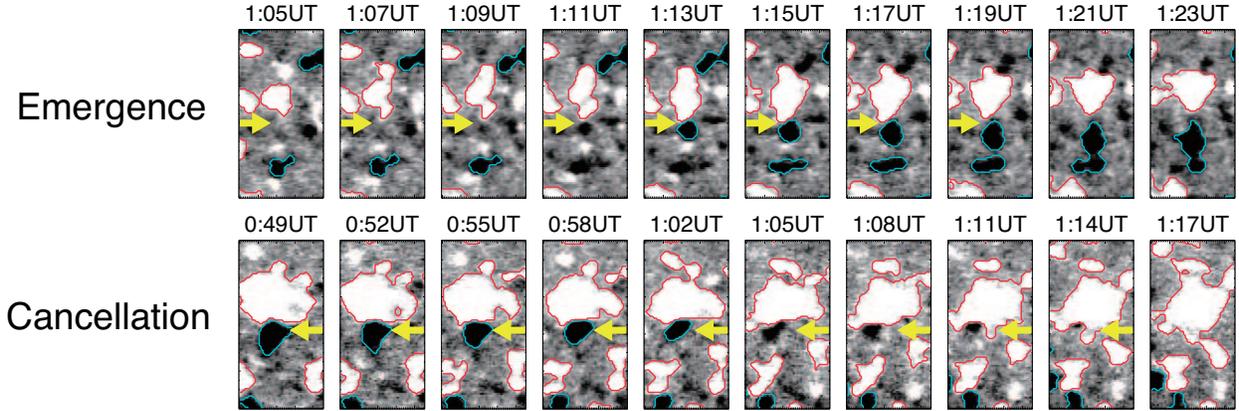}
\caption{Examples of detected emergences and cancellations. The background shows the magnetic flux density obtained by ${\it Hinode}$/NFI. The red and blue contours indicate positive and negative patches, respectively, detected with our threshold. The field of view is $7.2'' \times 14.4''$ for all the images.}
\end{figure}

\begin{figure}
\epsscale{0.99}
\plotone{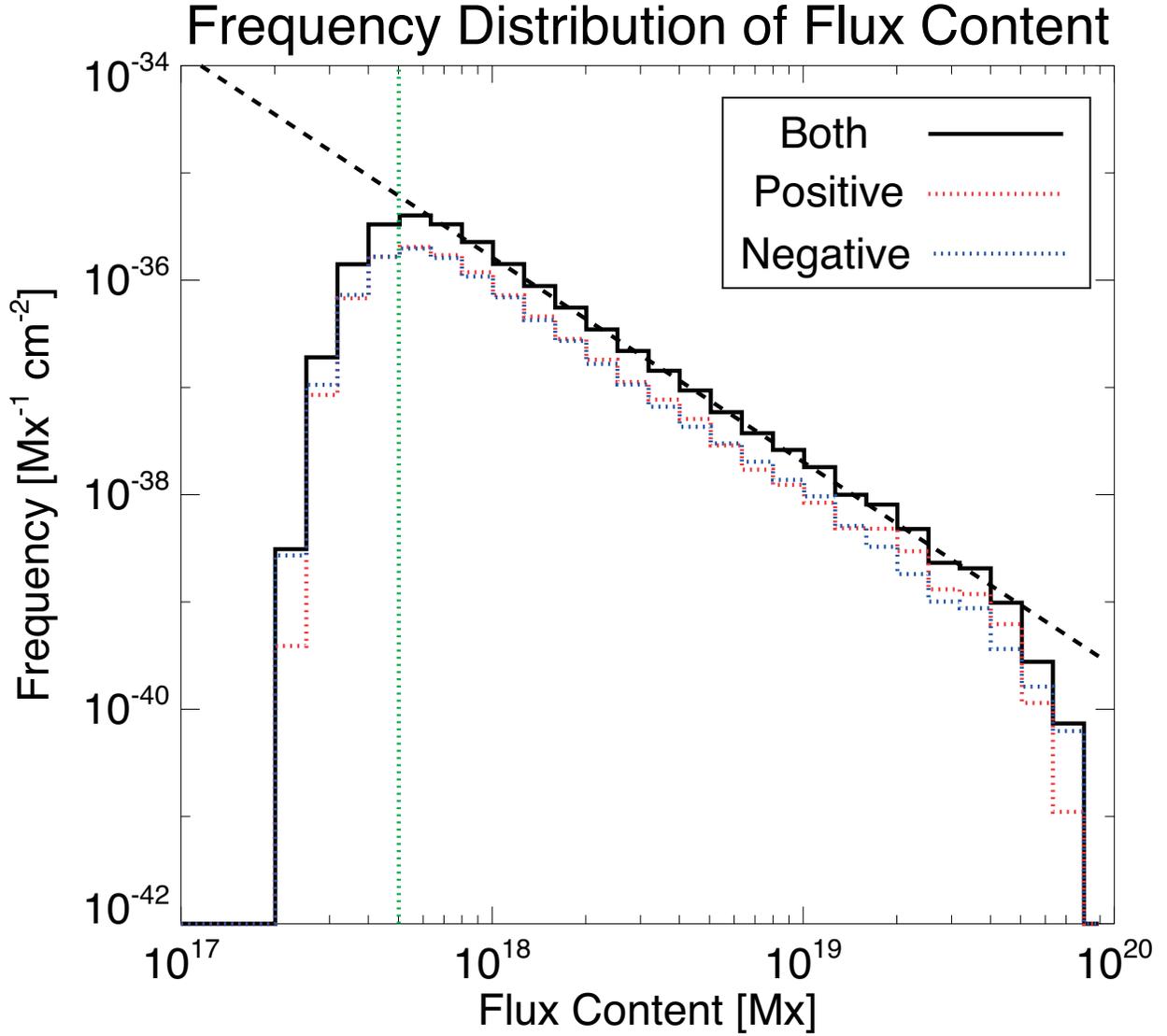}
\caption{Frequency distribution of the magnetic flux in Data2.
The red  and blue dashed histograms indicate the distributions of positive and negative polarities, respectively.
The solid histogram denotes the sum of these two distributions.
The black dashed line indicates the fitting result of the summed histogram.
The power-law index is $1.90$, with a fitting range between $10^{17.7}$ Mx and $10^{19}$ Mx.
The vertical green line denotes the detection limit, $\phi_{\rm th}=10^{17.7}$ Mx.}
\end{figure}

\begin{figure}
\epsscale{0.99}
\plotone{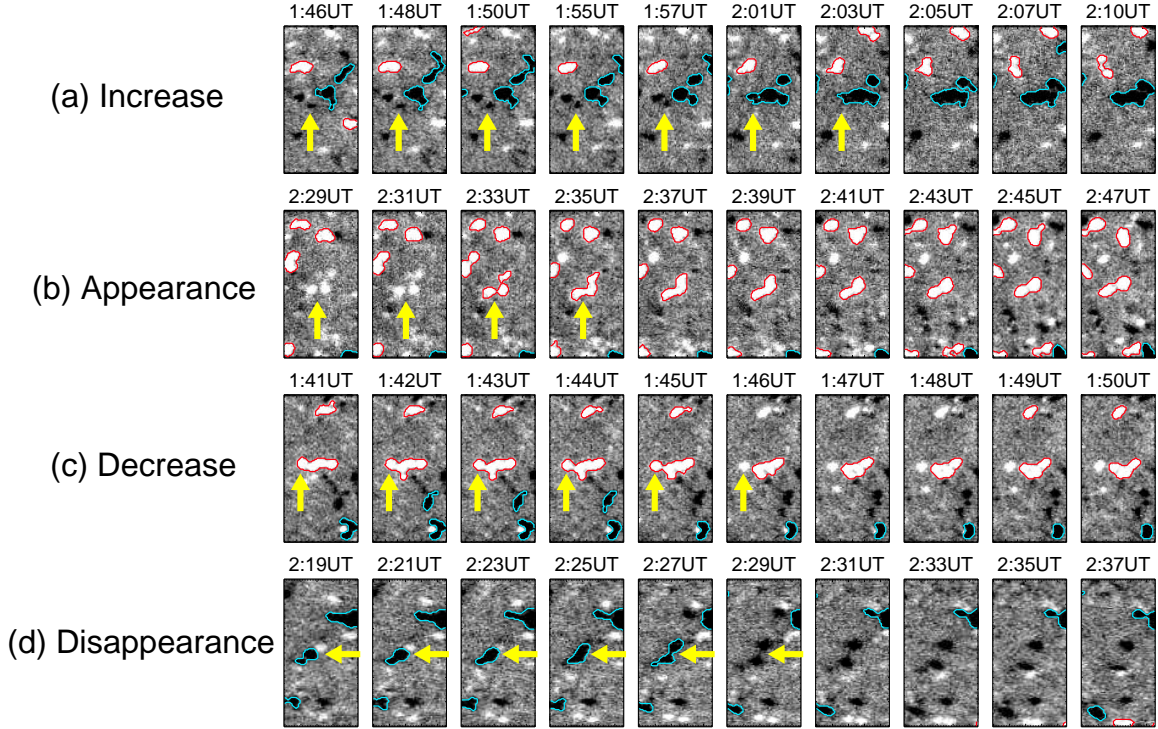}
\caption{Examples of apparent unipolar events in Data1, namely, (a) unipolar increase, (b) unipolar appearance, 
(c) unipolar decrease, and (d) unipolar disappearance. 
The background shows the magnetic flux density obtained by ${\it Hinode}$/NFI.
The red and blue contours indicate positive and negative patches, respectively, detected with our threshold.
The field of view is $9.6'' \times 19.2''$ for all the images.}
\end{figure}

\begin{figure}
\epsscale{0.99}
\plotone{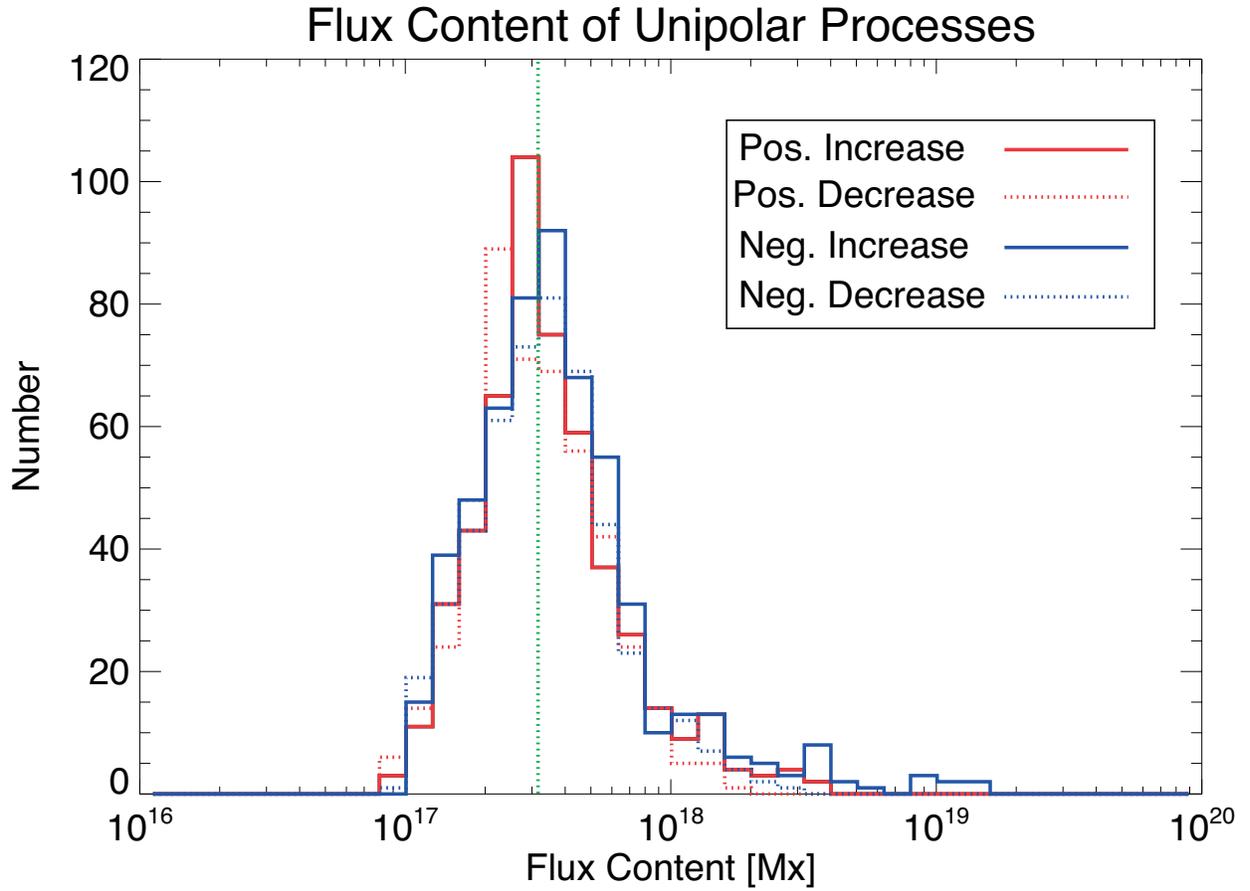}
\caption{Histogram of flux change during unipolar processes in Data1. The solid lines indicate increase processes, and the dotted lines indicate decrease processes. Colors indicate the polarity; red for positive and blue for negative. The vertical green line indicates the detection limit, $10^{17.5}$ Mx.}
\end{figure}

\begin{figure}
\epsscale{0.99}
\plotone{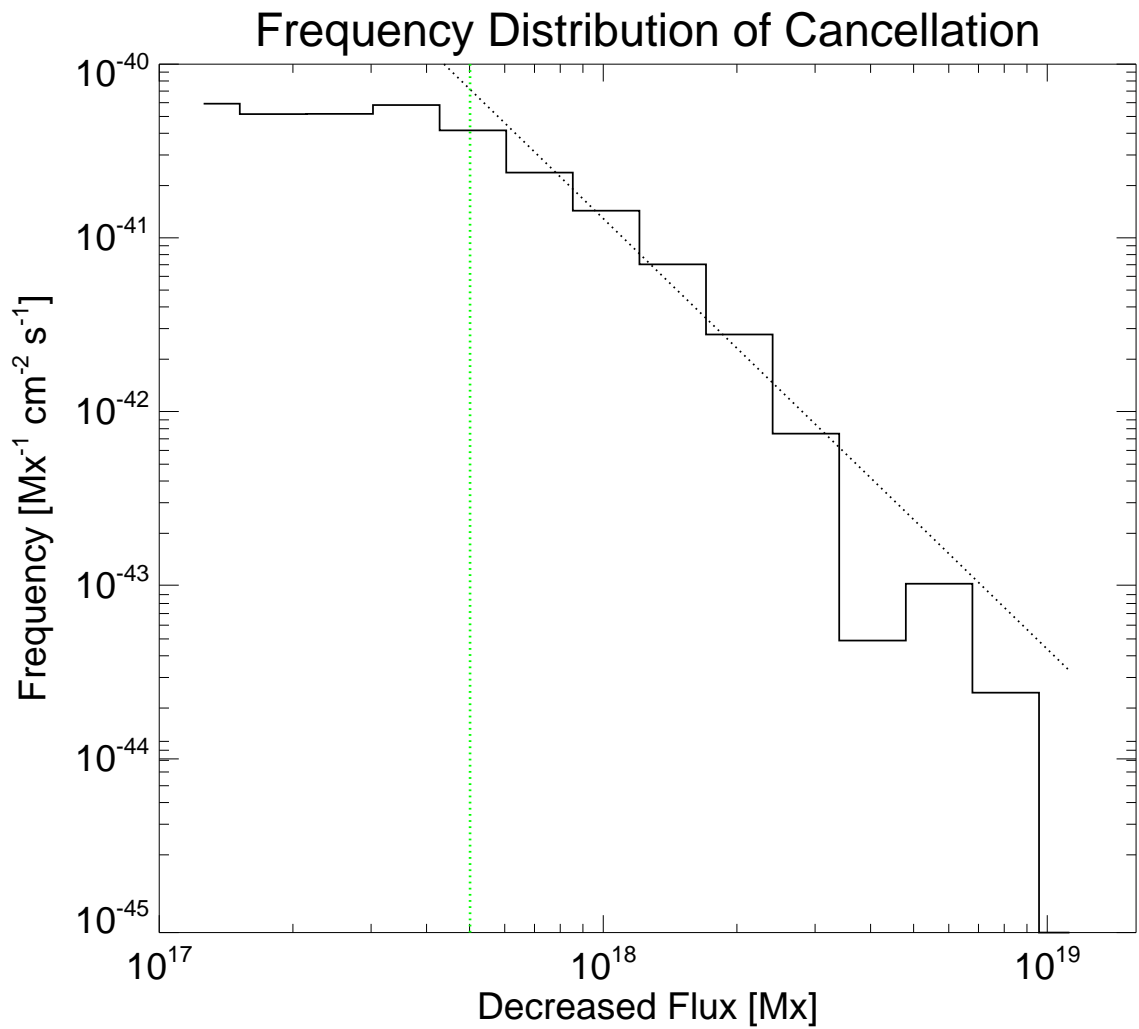}
\caption{Frequency distribution of decreased flux during cancellation with a binsize of 0.15 on log-scale. The histogram and dashed line indicate the observational result and fitting result with a power-law distribution, respectively. The fitting range is from $10^{17.7}$ Mx to $10^{18.4}$ Mx. The vertical green line shows the detection limit, $\phi_{\rm th}=10^{17.7}$ Mx. The fitted power-law index is $2.48 \pm 0.26$.}
\end{figure}

\begin{figure}
\epsscale{0.8}
\plotone{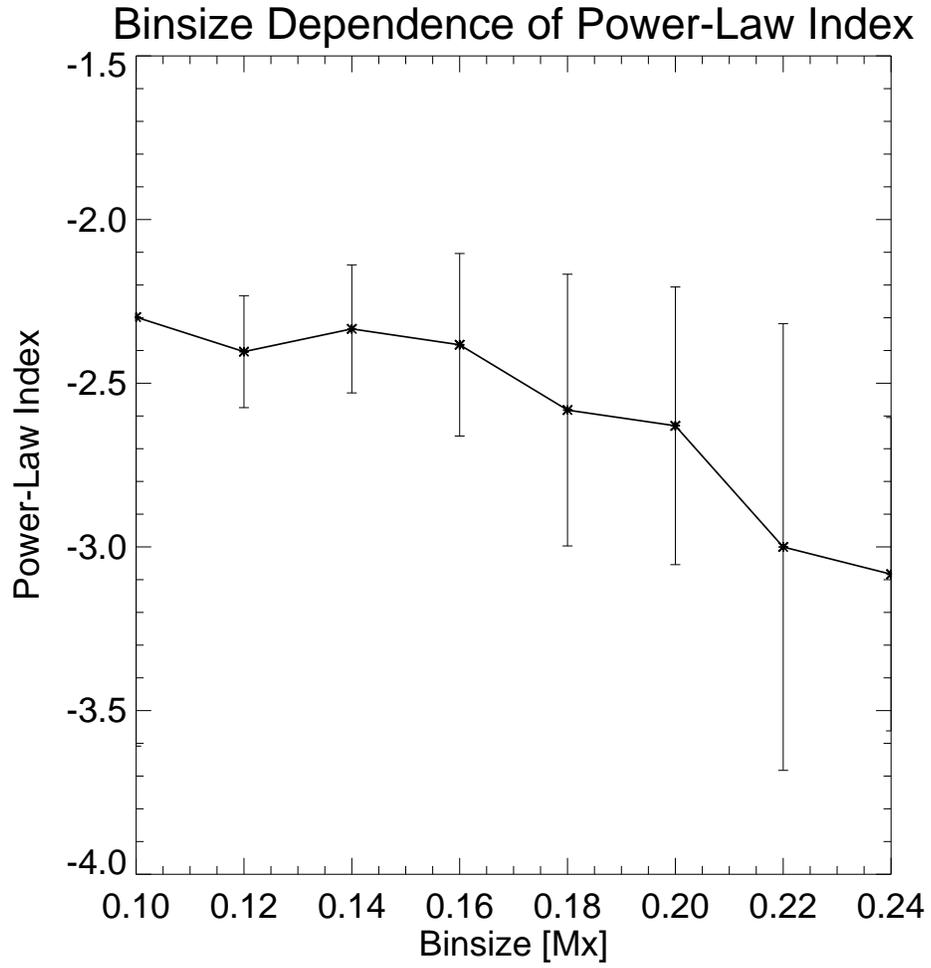}
\caption{The dependence of the obtained power-law index on binsize of the fitting. The bars shows 1$\sigma$-errors in each case.}
\end{figure}

\begin{figure}
\epsscale{0.9}
\plotone{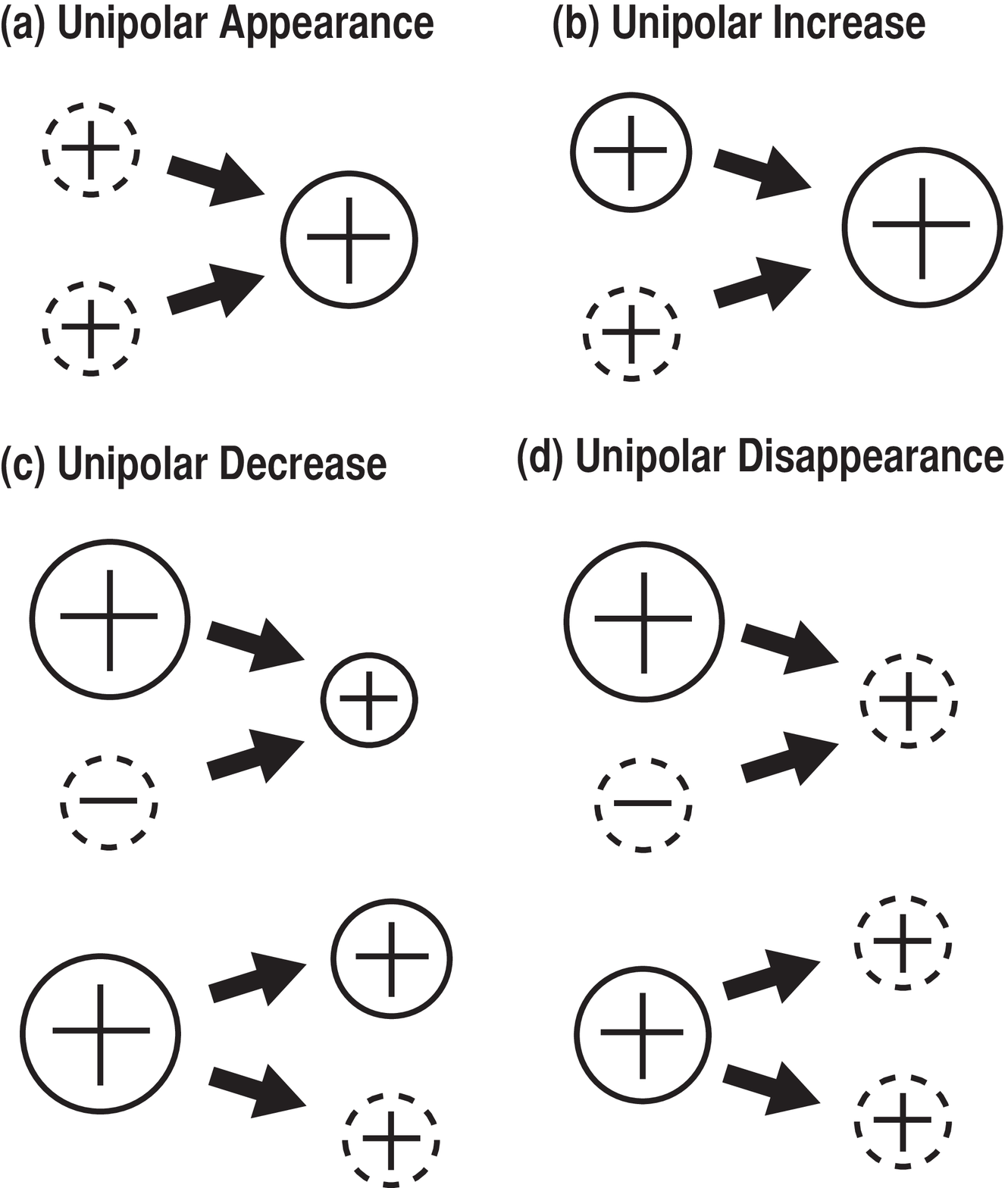}
\caption{Schematic of the interpretation of unipolar flux change events as interactions with and among patches below the detection limit.
(a) Unipolar appearance: Patches below the detection limit coalesce to one patch above the detection limit.
(b) Unipolar increase: Patch above the detection limit coalesces to a patch below the detection limit.
(c) Unipolar decrease: Patch above the detection limit cancels with a patch below the detection limit and stays above the detection limit, or patch above the detection limit splits into two patches, one of which is below the detection limit.
(d) Unipolar disappearance: Patch above the detection limit cancels with a patch below the detection limit and stays below the detection limit, or patch above the detection limit splits into patches, both of which are below the detection limit.}
\end{figure}

\begin{figure}
\epsscale{0.99}
\plotone{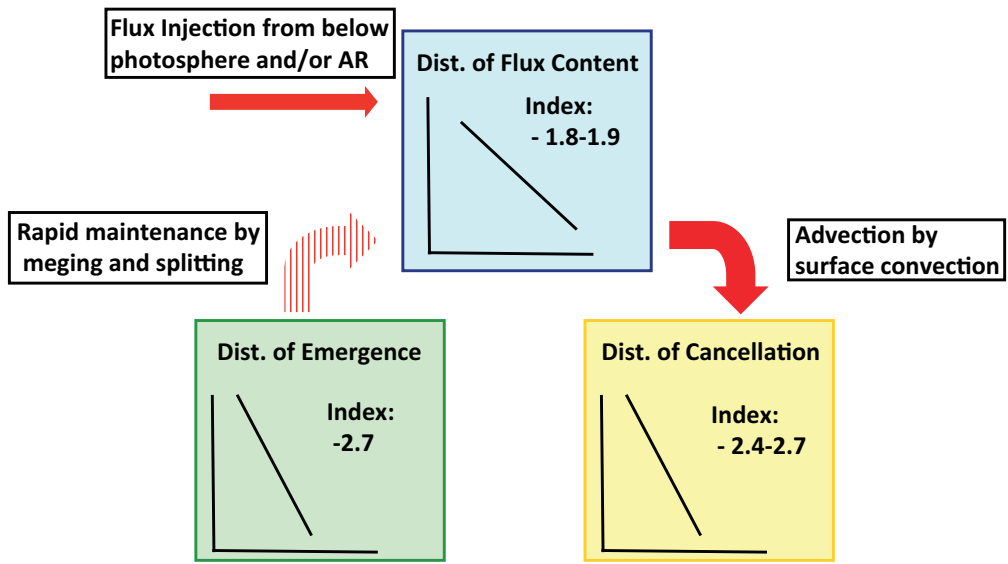}
\caption{Schematic picture of proposed scenario for flux maintenance in quiet regions. The frequency distribution of flux is maintained by rapid merging and splitting. The patches are advected by surface convection and accidental collisions between opposite polarities result in cancellations.}
\end{figure}

\begin{table}
\begin{tabular}{c|ccp{5zw}}
\hline
& Data1 & Data2 \\
\hline \hline
Period & \, & \, \\
\, Start & Nov. 11, 2009 0:30 UT & Dec. 30, 2008 10:24 UT \\
\, End & Nov. 11, 2009 4:09 UT & Jan. 5, 2009 5:37 UT \\
Duration & 3 h 39 min & 115 h 13 min \\
\hline
Center of FoV & \, &\, \\
\, Start & ($-14.1''$,$19.0''$)&($-520.9''$,$-9.3''$)\\
\, End & ($-20.2''$,$19.4''$)&($-695.9''$,$-6.9''$)\\
\hline
Pixel Size & $0.16''$ ($2 \times 2$ summed) & $0.16''$ ($2 \times 2$ summed) \\
Temporal Cadence & $\thicksim$1 min & $\thicksim$5 min \\
Number of Images & 199 & 1642 \\
FoV & $113'' \times 113''$ & $94.9'' \times 91.4''$ \\
Observation Mode & Stokes IV+DG & Stokes V/I \\
Wavelength Offset & +/-160m\AA & +140m\AA \\
\hline \hline 
\end{tabular}
\caption{Summary of observational parameters.}
\end{table}

\begin{table}
\begin{tabular}{c|cc|cc}
\hline 
             &\multicolumn{2}{c|}{Data1}              &\multicolumn{2}{c}{Data2} \\
             &Positive           &Negative           &Positive           &Negative \\ \hline \hline 
Tracked Patches &1636               &1637               &21823              &19544               \\ 
             &(2.53 Mx cm$^{-2}$)&(3.60 Mx cm$^{-2}$)&(3.76 Mx cm$^{-2}$)&(3.42 Mx cm$^{-2}$) \\ \hline
Splitting    &536                &535                &5905               &4252 \\
Merging      &493                &482                &5764               &4143 \\ \hline
Emergence    &\multicolumn{2}{c|}{3}                  &\multicolumn{2}{c}{27} \\ 
Cancellation &\multicolumn{2}{c|}{86}                 &\multicolumn{2}{c}{775} \\ \hline
Unpaired Increase&503                &556                &6185               &7633 \\ 
Unpaired Decrease&381                &398                &5440               &5400 \\ \hline
\end{tabular}
\caption{Summary of number of detected patches and processes in Data1 and Data2.}
\end{table}

\end{document}